\documentclass[aps,prl,reprint,superscriptaddress,floatfix]{revtex4-2}
\usepackage{graphicx}  
\usepackage{amsmath}   
\usepackage{amssymb}   
\usepackage{bm}        
\usepackage{hyperref}  
\usepackage{physics}
\usepackage{tikz-cd}
\usepackage{appendix}
\hypersetup{colorlinks=true, urlcolor=blue, citecolor=blue}

\begin{document}

\title{Topology Does Not Determine Defect Mobility in Active Polar Matter}
\author{Abhra Puitandy}
\email{abhrapuitandy.rs.phy22@itbhu.ac.in}
\affiliation{Department of Physics,Indian Institute of Technology(BHU), Varanasi-221005,India}
\author{Shradha Mishra}
\email{smishra.phy@iitbhu.ac.in}
\affiliation{Department of Physics,Indian Institute of Technology(BHU), Varanasi-221005,India}


\begin{abstract}
Topological defects are commonly characterized by their charge, although defects with the same charge can have different internal morphologies. We study the annihilation of isolated $\pm1$ defect pairs in a two-dimensional compressible Toner--Tu model, varying the morphology of the $+1$ defect while keeping the initial charges and separation fixed. Activity produces a morphology-dependent mobility asymmetry: reversing the radial orientation of an aster reverses which defect travels farther, and asters show the strongest response. An early-time analysis reveals that the initial morphology selects distinct density modes whose gradients subsequently feed back on the polarization. Consistently, removing the density-gradient coupling strongly suppresses the asymmetry. These results establish defect morphology, in addition to topological charge, as a dynamical variable governing defect-pair motion in active polar matter.
\end{abstract}

\maketitle

Active matter consists of interacting units that continuously consume energy and convert it into directed motion, generating collective behavior far from thermodynamic equilibrium \cite{fodor2016far}. Examples range from bacterial suspensions \cite{aranson2022bacterial,sokolov2012physical,jena2025spatio} and cytoskeletal assemblies \cite{needleman2017active,banerjee2020actin,sanchez2012spontaneous} to bird flocks \cite{cavagna2014bird} and synthetic active colloids \cite{zottl2016emergent,bishop2023active,lefranc2025synthetic,das2018ordering}. The Toner--Tu theory provides a paradigmatic framework for collective motion in dry polar systems and predicts long-range orientational order and giant density fluctuations in two dimensions \cite{toner1995long,toner1998flocks}.

Topological defects are prominent structures in ordered active media. Their motion, interactions, and annihilation are coupled to the nonequilibrium dynamics of the surrounding medium \cite{peng2016command,shankar2022topological}. This coupling is well studied in active nematics, where the dynamics of $\pm1/2$ defects underlies defect proliferation, active turbulence, and transport by coherent structures \cite{giomi2013defect,giomi2014defect,alert2022active,serra2023defect,sarathi2026spatiotemporal,mishra2025string}. By contrast, the dynamics of full-integer defects in dry polar active matter is less understood, even though defects with the same charge can have distinct polarization textures.

In active matter, a defect's internal polarization structure can shape the local density and polarization fields and thereby affect its motion even when its topological charge is unchanged. This issue is particularly relevant for $+1$ defects, which can form asters, spirals, or vortices \cite{kruse2004asters,ronning2023spontaneous,mondal2025coarsening}. Density--polarization coupling can in turn influence the orientational structure of active polar systems \cite{dedenon2023density,mondal2025coarsening}, and recent work has highlighted the role of local defect structure in active defect motion \cite{la2026shape}. These observations motivate a central question: does topological charge alone characterize defect dynamics, or can morphology provide an additional dynamical degree of freedom?

Here we address this question using a two-dimensional compressible Toner--Tu model of dry polar active matter. We prepare isolated $\pm1$ pairs at the same initial separation and vary the morphology of the $+1$ defect among converging and diverging asters, spirals, and a vortex. Activity produces a pronounced mobility asymmetry, including a reversal of which defect travels farther when the radial orientation of the $+1$ texture is reversed. An early-time analysis shows that morphology selects distinct density modes, whose gradients subsequently feed back on the polarization; the asymmetry is strongly suppressed when this density-gradient coupling is removed. Our results identify defect morphology as an additional dynamical variable, beyond topological charge, that influences defect motion and annihilation in active polar matter.

\textit{The Model--}
We consider a two-dimensional dry compressible polar system described by
the Toner--Tu equations for the density $\rho(\mathbf r,t)$ and polarization
$\mathbf p(\mathbf r,t)$,
\begin{align}
\partial_t\rho+v_0\nabla\cdot(\rho\mathbf p)&=0,
\label{eq:density}\\
\partial_t\mathbf p+\lambda_1(\mathbf p\cdot\nabla)\mathbf p
&=(\alpha-\beta|\mathbf p|^2)\mathbf p
-\frac{v_0\sigma}{2\rho_0}\nabla\rho\notag \\
&+\lambda_2\mathbf p(\nabla\cdot\mathbf p)
+\lambda_3 p_i\nabla p_i
+D_b\nabla^2\mathbf p.
\label{eq:polarization}
\end{align}
here, $v_0$ controls the activity, while $\sigma$ sets the strength of the density-gradient contribution to the polarization dynamics. The remaining parameters are $\alpha=0.25$, $\beta=1$, $D_b=1$, and $\rho_0=0.75$. Following \cite{mishra2010fluctuations}, we choose $\lambda_1=0.36v_0$ and $\lambda_2=\lambda_3=0.12v_0$. We vary $v_0$ and $\sigma$ to determine how activity and density--polarization coupling affect defect-pair dynamics.

\begin{figure}
    \centering
    \includegraphics[width=0.95\linewidth]{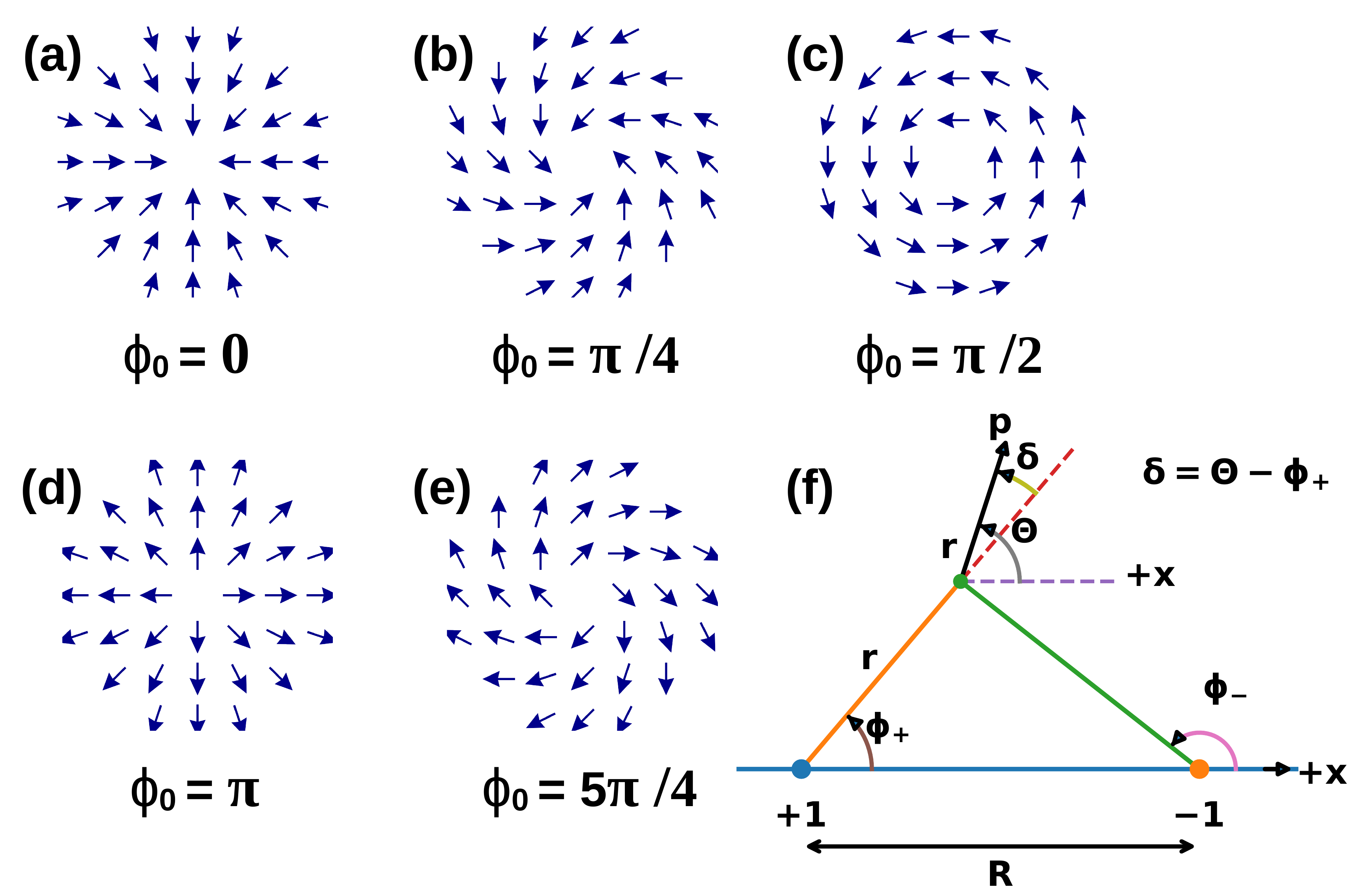}
    \caption{Polarization field line patterns around the different types of $+1$ defect: (a) Converging aster (CA) (b) Converging spiral (CS) (c) Vortex (VR) (d) Diverging aster (DA) (e) Diverging spiral (DS) , studied in this paper along with their corresponding phase offsets $\phi_0$. Panel (f) shows a $+1/-1$ defect pair used in the analytical calculation is shown here. The $+1$ defect defines the local origin and the defect-pair axis defines the $x$ direction, with the $-1$ defect a distance $R$ away.
A point $\mathbf{r}$ near the $+1$ defect is specified by $(r,\phi)$,
while $\phi_{-}$ is its polar angle relative to the $-1$ defect.
The polarization orientation is $\Theta$, and $\delta=\Theta-\phi$ is the angle between the polarization and the
local radial direction. }
    \label{fig:deftype.pdf}
\end{figure}

We initialize a single $\pm1$ pair at $\mathbf r_+=(L_x/4,L_y/2)$ and $\mathbf r_-=(3L_x/4,L_y/2)$, so that all realizations begin with the same defect separation. Where $L_x$ and $L_y$ are the size of the box in $x$ and $y$ directions respectively. The polarization angle is initialized as
\begin{equation}
\Theta(\mathbf r)=
\arg(\mathbf r-\mathbf r_+)
-\arg(\mathbf r-\mathbf r_-)-\phi_0 ,
\label{eq:initial_angle}
\end{equation}
where the offset $\phi_0$ changes the polarization morphology without changing the defect charges or their initial positions.  We use $\phi_0=0,\pi/4,\pi/2, \pi$ and $ 5\pi/4,$  to generate, respectively, a converging aster (CA), converging spiral (CS), vortex (VR), diverging aster (DA), and diverging spiral (DS). The corresponding polarization textures near the defect core are shown in Fig.~\ref{fig:deftype.pdf}(a-e) respectively. The density is initially uniform, $\rho(\mathbf r,0)=\rho_0$. This construction therefore allows us to vary the internal morphology of the $+1$ defect while keeping the topology and initial pair geometry fixed. 

Equations~(\ref{eq:density}) and (\ref{eq:polarization}) are integrated on a $256\times256$ periodic domain.
 Defects are located from the winding of the polarization angle around elementary plaquettes. 
Further details of the numerical implementation, defect tracking, and robustness checks are given
in the Supplemental Material.\\

\begin{figure*}
\centering
\includegraphics[width=1.0\linewidth]{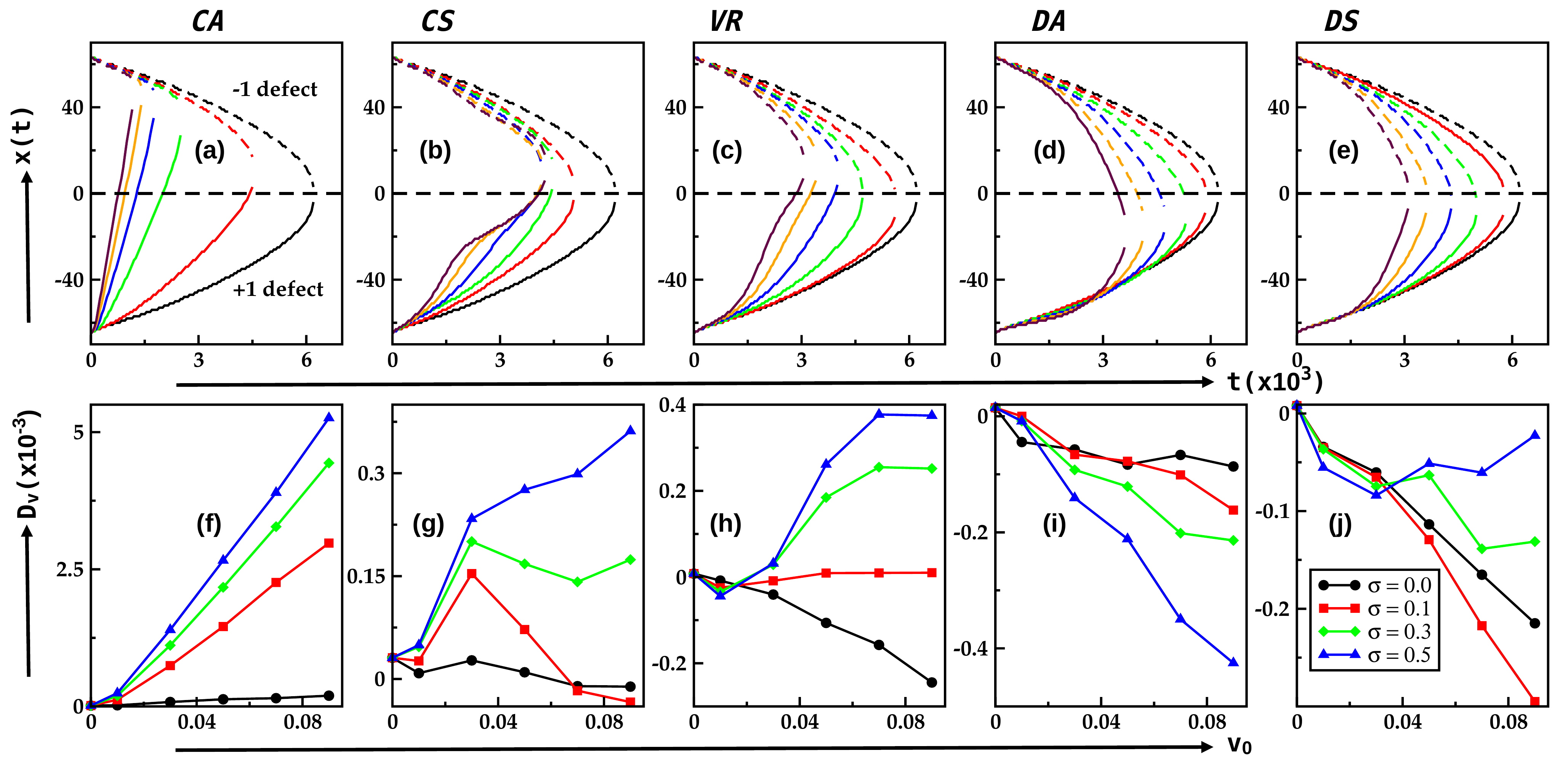}
\caption{ Positions $x(t)$ of the $+1$ (solid lines) and $-1$ (dashed lines) defects
during annihilation at $\sigma=0.5$ for initial $+1$ morphologies:
(a) vortex, (b) converging spiral, (c) diverging spiral,
(d) converging aster, and (e) diverging aster.
Colors denote the activity $v_0$. Panels (f)-(j) shows velocity difference,
$D_v = v_{+1}-v_{-1}$,
between the $+1$ and $-1$ defects as a function of activity $v_0$. Different symbols and colours denote
$\sigma=0.0$, $0.1$, $0.3$, and $0.5$. }
\label{fig:def_traj4.pdf}
\end{figure*}

\textit{Morphology-dependent defect mobility.--}
We first examine how the internal structure of the $+1$ defect influences
the annihilation dynamics. Figure~\ref{fig:def_traj4.pdf}(a-e) shows the time trajectory of the  $+1$ and
$-1$ defects during annihilation for the five initial morphologies at
$\sigma=0.5$. In the passive limit, $v_0=0$, the two defects approach each
other nearly symmetrically for all morphologies. Finite activity breaks
this near-symmetric motion, and the two defects no longer travel
comparable distances before annihilation. The resulting asymmetry depends
strongly on the initial $+1$ morphology.

The clearest contrast occurs between the two asters. For a converging
aster, increasing activity enhances the motion of the $+1$ defect relative
to its $-1$ partner [see SM Movie 1.1], whereas for a diverging aster the opposite occurs and
the $-1$ defect travels farther [see SM Movie 1.4]. The corresponding spirals show the same
converging--diverging distinction, but with a weaker separation between
the two trajectories [see SM Movie 1.2 and Movie 1.5]. The vortex exhibits a weaker and nonmonotonic
response: at low activity the $-1$ defect can travel farther, whereas at
larger activity the relative motion reverses and the $+1$ defect becomes
more mobile [see SM Movie 1.3]. Thus, changing the polarization texture of the $+1$ defect,
without changing its charge or the initial pair geometry, changes both
the degree of asymmetry in the trajectories and which defect travels
farther before annihilation.\\
To quantify the unequal motion of the two defects, we define an effective
migration speed from the net forward distance traveled before annihilation.
Denoting the resulting forward
distances by $R_+$ and $R_-$, we define the mean velocity of $\pm$ defects,
\begin{equation}
V_{\pm}=\frac{R_{\pm}}{t_{\rm ann}},
\qquad
D_v=V_+-V_-,
\label{eq:Dv}
\end{equation}

where $t_{\rm ann}$ is the annihilation time. Thus, $D_v>0$ indicates larger
effective migration of the $+1$ defect, whereas $D_v<0$ indicates larger
migration of the $-1$ defect. Fig.~\ref{fig:def_traj4.pdf}(f-j) shows $D_v$ as a function of activity averaged over five initial
morphologies and different values of $\sigma$. 
At $v_0=0$, $D_v$ remains close to zero for all morphologies, consistent with the nearly symmetric trajectories in Fig.~\ref{fig:def_traj4.pdf}(a--e). Finite activity produces a pronounced morphology dependence. Most notably, the converging and diverging asters exhibit opposite signs of $D_v$: $D_v>0$ for the former and $D_v<0$ for the latter. Thus, reversing the radial orientation of a $+1$ texture reverses the relative migration of the corresponding $+1/-1$ pair. The spirals exhibit the same sign distinction, but with a much smaller magnitude, whereas the vortex shows a weaker, nonmonotonic response, with $D_v$ changing sign as activity increases.

The dependence on $\sigma$ further highlights the role of density--polarization coupling. For $\sigma=0$, the mobility difference remains small even at finite activity, whereas finite $\sigma$ produces a substantially larger $|D_v|$, particularly for the asters. Since $\sigma$ controls the coupling of density gradients to the polarization field in Eq.(\ref{eq:polarization}), this contrast suggests that density--polarization feedback contributes significantly to the observed mobility asymmetry.



\textit{Density--polarization feedback.--} To address this question, we examine how the different initial morphologies
generate distinct early-time density and polarization responses.
The
calculation is performed near the $(+1)$ defect while retaining the leading
distortion produced by its $(-1)$ partner.
Taking the $+1$ defect as the local origin and the
defect-pair axis along the $x$ direction, we place the $-1$ defect at
$R\hat{\mathbf{x}}$, where $R$ is the defect separation (as shown in Fig.\ref{fig:deftype.pdf}(f). A point close
to the $+1$ core is written as,
$\mathbf{r}=r(\cos\phi,\sin\phi)$, and 
$r\ll R$. The angular field of the seeded defect pair Eq.(\ref{eq:initial_angle}) can then be
written as, $\Theta(\mathbf{r})=\phi_{+}-\phi_{-}-\phi_{0}$, where $\phi_{+}$ and $\phi_{-}$ are the polar angles measured relative to the $+1$ and $-1$ defect cores, respectively as shown in Fig.\ref{fig:deftype.pdf}(f). Since the observation
point is described relative to the $+1$ core, $\phi_{+}=\phi$, while
the angle relative to the $-1$ defect is
\[
\phi_-=\pi-\tan^{-1}\!\left(
\frac{r\sin\phi}{r\cos\phi-R}
\right).
\]

For $r/R\ll1$, expansion about the $+1$ core gives
\begin{equation}
\phi_{-}
=\pi-\frac{r}{R}\sin\phi
+\mathcal{O}\!\left(\frac{r^{2}}{R^{2}}\right).
\end{equation}
Substituting this result into Eq.(\ref{eq:initial_angle}), we obtain
\begin{equation}
\Theta(r,\phi)
=
\phi-\pi-\phi_{0}
+\frac{r}{R}\sin\phi
+\mathcal{O}\!\left(\frac{r^{2}}{R^{2}}\right).
\label{eq:pairdistort}
\end{equation}
The first three terms describe the locally axisymmetric $+1$ texture, whereas the term proportional to $r/R$ is the leading distortion
produced by the neighboring $-1$ defect. Here $\phi$ is measured from the defect-pair axis. 
Since the density is initially uniform, $\rho(\mathbf{r},0)=\rho_0$,
Eq.(\ref{eq:density}) gives
$\left.\partial_t\rho\right|_0=-v_0\rho_0\nabla\cdot\mathbf{p}$.
Writing the polarization field as
$\mathbf{p}=f(r)(\cos\Theta,  
 \sin\Theta)$, where $f(r)$ denotes its
radial magnitude profile, and using Eq.(\ref{eq:pairdistort}), we obtain, to leading
order in $r/R$,
\begin{align}
\nabla\cdot\mathbf{p}
=
-\cos\phi_0\left(f'+\frac{f}{r}\right)
-\frac{f}{R}\cos\phi_0\cos\phi \notag \\
-\frac{2f+rf'}{R}\sin\phi_0\sin\phi
+\mathcal{O}\!\left(\frac{r^2}{R^2}\right),
\end{align}
where $f'=df/dr$. The initial density response therefore separates
naturally into an isotropic contribution and longitudinal and
transverse first angular harmonics,
\begin{equation}
\left.\partial_t\rho\right|_0
=
D_0(r)+D_c(r)\cos\phi+D_s(r)\sin\phi .
\label{eq:density_modes}
\end{equation}
The corresponding mode amplitudes are
$D_0=v_0\rho_0\cos\phi_0(f'+\frac{f}{r})$,
$D_c=(v_0\rho_0\frac{f}{R})\cos\phi_0$, and
$D_s=\frac{v_0\rho_0}{R}(2f+rf')\sin\phi_0$.
Here $D_0$ gives the angle-independent rate of density change,
$D_c\cos\phi$ describes the longitudinal front--back variation along
the defect-pair axis, and $D_s\sin\phi$ describes the transverse
variation. The detailed expressions for the mode amplitudes are given in the \textcolor{red}{SM(sec.A1)}.\\ 
\begin{table}[t]
\centering
\caption{
Initial density- and polarization-rate modes for the five $+1$-defect
morphologies. For the density modes, $+$ and $-$ denote the sign of the
corresponding mode amplitude, while $0$ denotes a mode that vanishes at
leading order. For the polarization modes, $\checkmark$ denotes a mode
that is generally nonzero and $0$ a mode that vanishes identically by
symmetry. CA, CS, V, DS, and DA denote the converging aster, converging
spiral, vortex, diverging spiral, and diverging aster, respectively.
}
\label{tab:mode_summary}
\setlength{\tabcolsep}{1.0pt}
\begin{tabular}{c|ccc|cccccc}
\hline\hline
&
\multicolumn{3}{c|}{Density-rate modes}
&
\multicolumn{6}{c}{Polarization-rate modes}
\\
Morphology
& $D_0$ &   $D_c$ & $D_s$
& $P_0^{(r)}$ & $P_0^{(\phi)}$
& $P_c^{(r)}$ & $P_c^{(\phi)}$
& $P_s^{(r)}$ & $P_s^{(\phi)}$
\\
\hline
CA
& $+$ & $+$ & $0$
& $\checkmark$ & $0$ & $\checkmark$ & $0$ & $0$ & $\checkmark$
\\
CS
& $+$ & $+$ & $+$
& $\checkmark$ & $\checkmark$ & $\checkmark$ & $\checkmark$
& $\checkmark$ & $\checkmark$
\\
VR
& $0$ & $0$ & $+$
& $\checkmark$ & $\checkmark$ & $\checkmark$ & $\checkmark$
& $\checkmark$ & $\checkmark$
\\
DA
& $-$ & $-$ & $0$
& $\checkmark$ & $0$ & $\checkmark$ & $0$ & $0$ & $\checkmark$
\\
DS
& $-$ & $-$ & $-$
& $\checkmark$ & $\checkmark$ & $\checkmark$ & $\checkmark$
& $\checkmark$ & $\checkmark$
\\
\hline\hline
\end{tabular}
\end{table}

Table~\ref{tab:mode_summary} summarizes the resulting density-mode hierarchy. For the converging aster, $D_0,D_c>0$ and $D_s=0$: the density starts increasing around the core and develops a front--back difference along the pair axis. Reversing the radial polarization reverses both nonzero
modes, giving $D_0,D_c<0$ and $D_s=0$ for the diverging aster. The spirals show the same reversal between their converging and diverging forms, but their isotropic and longitudinal rates are smaller than for the corresponding asters, and a transverse contribution is also present. For the vortex, $D_0=D_c=0$, so its initial density change is purely transverse.\\
This result provides a direct connection between morphology and the density field that subsequently acts on the polarization. Initially $\nabla\rho=0$, so the density-gradient term in the polarization equation vanishes. However, after a short time,
\begin{equation}
\rho(\mathbf r,t)
=\rho_0+t\left.\partial_t\rho\right|_0+O(t^2),
\label{eq:rho_short}
\end{equation}
the morphology-dependent density changes in Eq.~(\ref{eq:density_modes})
generate different density gradients as the system begins to evolve.\\
Notably, the longitudinal density-rate mode has one sign for the
converging textures, the opposite sign for the diverging textures, and
vanishes for the vortex. This ordering follows the observed change in the
sign of $D_v$ across the five morphologies. This correspondence does not by itself determine the defect velocity. However, together with the
strong suppression of $D_v$ at $\sigma=0$, it points to density--polarization feedback as an important source of the morphology-dependent motion. Since topological defects are singularities of the polarization field, their motion is inherently linked to the spatiotemporal evolution of the surrounding polarization texture. It is therefore important to examine how the different initial defect morphologies evolve under the polarization dynamics.\\
\textit{Morphological evolution.--}
The polarization texture that generates this
density change need not retain its initial form. To characterize its deformation, we resolve the polarization into components radial and azimuthal to the $+1$ defect core, $\mathbf{p}=p_r\mathbf{e}_r+p_\phi\mathbf{e}_\phi$, with $p_r=f(r)\cos\delta$, $p_\phi=f(r)\sin\delta$, and $\delta=\Theta-\phi$. From Eq.(\ref{eq:pairdistort}), $\delta=-\pi-\phi_0 +(r/R)\sin\phi+\mathcal{O}(r^2/R^2)$. Defining $A=-\cos\phi_0$ and $B=\sin\phi_0$, expansion to first order in $r/R$ gives
\begin{align}
p_r=f(r)\left[A-B\frac{r}{R}\sin\phi\right]
+\mathcal{O}\!\left(\frac{r^2}{R^2}\right)
\label{eq:pr}\\
p_\phi=f(r)\left[B+A\frac{r}{R}\sin\phi\right]
+\mathcal{O}\!\left(\frac{r^2}{R^2}\right).
\label{eq:pphi}
\end{align}
At the initial instant $\nabla\rho=0$, so the density-gradient term in Eq.(\ref{eq:polarization}) does not yet contribute to the polarization rate. Substituting the above fields into Eq.(\ref{eq:polarization}) and consistently retaining terms through first order in $r/R$, the radial and azimuthal polarization rates decompose into the same three angular harmonics,
\begin{align}
\left.\partial_t p_r\right|_0
&=P^{(r)}_0(r)+P^{(r)}_c(r)\cos\phi
+P^{(r)}_s(r)\sin\phi,\\
\left.\partial_t p_\phi\right|_0
&=P^{(\phi)}_0(r)+P^{(\phi)}_c(r)\cos\phi
+P^{(\phi)}_s(r)\sin\phi.
\label{eq:pol_modes}
\end{align}

The explicit mode amplitudes are derived in
\textcolor{red}{SM(sectionA2)(g)}, and their symmetry structure
is summarized in Table~\ref{tab:mode_summary}. The radial polarization
modes describe changes in the inward or outward
component of the texture. Specifically,
$P_0^{(r)}$ represents an angularly uniform
change in the signed radial component, whereas
$P_c^{(r)}$ produces a front--back difference
along the defect-pair axis and $P_s^{(r)}$
produces an upper--lower difference across it.
The corresponding azimuthal modes describe
changes in the circulating component:
$P_0^{(\phi)}$ modifies it uniformly around
the core, while $P_c^{(\phi)}$ and
$P_s^{(\phi)}$ generate longitudinal and
transverse asymmetries, respectively.
Thus, the radial and azimuthal modes distinguish
changes in the two polarization components,
while their angular harmonics identify the
spatial symmetry of those changes.\\
For both converging and diverging asters,
$P_s^{(r)}=P_0^{(\phi)}=P_c^{(\phi)}=0$,
while $P_0^{(r)}$, $P_c^{(r)}$, and
$P_s^{(\phi)}$ are generally nonzero.
Their radial component can therefore change
uniformly and develop a front--back asymmetry,
while the neighboring defect can induce an
upper--lower asymmetric azimuthal bending.
However, the absence of an isotropic azimuthal
mode implies that the angularly symmetric
part of the texture does not initially acquire
a circulating component. Although the two
asters obey the same symmetry restrictions,
their mode amplitudes differ because their
radial orientations are opposite.\\
For the converging and diverging spirals,
all six modes are generally allowed.
Consequently, both radial and azimuthal
components can evolve through uniform,
longitudinal, and transverse responses,
allowing changes in their relative strengths
and hence in the local spiral geometry.
The corresponding mode amplitudes differ
between the two spirals, so their initial
deformations need not proceed in the same
direction. The vortex also permits all six
modes. In particular, its nonzero isotropic
radial mode $P_0^{(r)}$ shows that an
initially tangential texture can acquire
a radial component, thereby departing
from its purely vortex-like form.\\
\begin{figure}
\centering
\includegraphics[width=1.0\linewidth]{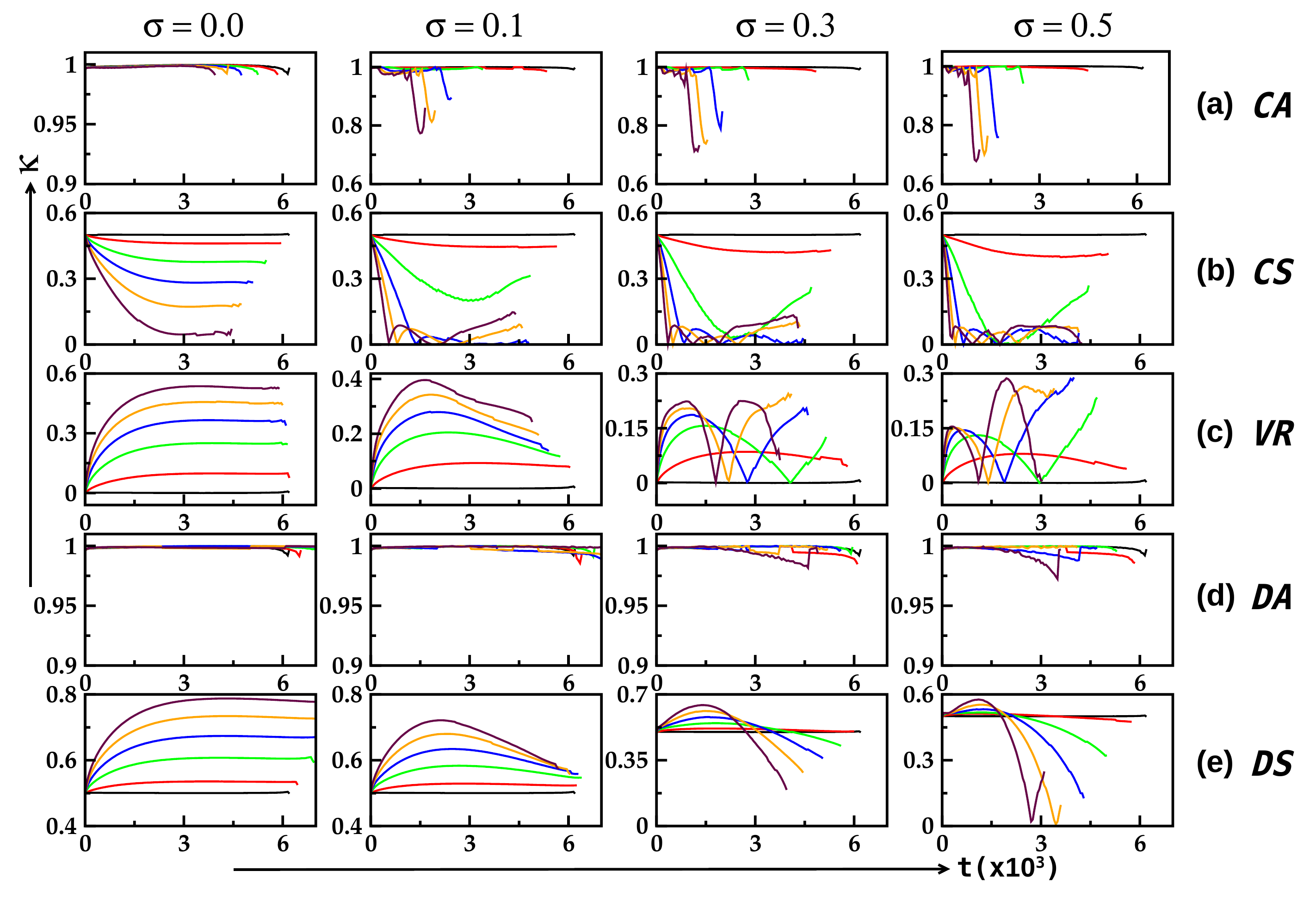}
\caption{Time evolution of the morphology parameter $\kappa$ for different initial integer-defect morphologies and model parameters. Rows correspond to vortex(VR), converging spiral (CS), diverging spiral (DS), converging aster (CA), and diverging aster (DA) initial conditions. Columns correspond to compressibility values $\sigma=0.0$, $0.1$, $0.3$, and $0.5$. Different colors denote activity strengths $v=0.0$ (black), $0.01$ (red), $0.03$ (green), $0.05$ (blue), $0.07$ (orange), and $0.09$ (maroon). The morphology parameter is defined as
$\kappa=\frac{|S|}{|S|+|\Omega|}$,
where $S=\nabla\cdot\mathbf{p}$ and $\Omega=(\nabla\times\mathbf{p})z$ are evaluated within the defect-core region. }
\label{fig:shape3.pdf}
\end{figure}
The polarization modes therefore identify
the symmetry and spatial character of the
initial changes in the texture. However,
they do not by themselves establish whether
the resulting defect becomes more radial
or more circulating.

To determine what these changes mean for the defect morphology itself, we define a deformation parameter $\kappa$.
\begin{equation}
\kappa=
\frac{|S|}{|S|+|\Omega|},
\qquad
S=\nabla\cdot\mathbf p,
\qquad
\Omega=(\nabla\times\mathbf p)_z,
\label{eq:kappa}
\end{equation}
Here $\kappa=1$ corresponds to a purely radial aster, whereas $\kappa=0$ corresponds to a purely azimuthal vortex. To determine its early-time evolution, we introduce the local orientation
$\delta=\Theta-\phi$. Using Eq (\ref{eq:pairdistort}), we evaluate the splay and
vorticity on a symmetric contour at the core scale $r=r_c$. The $\mathcal{O}(r_c/R)$ corrections are first angular harmonics and therefore vanish upon angular averaging, giving
\begin{align}
\langle S\rangle_\phi
=
\left(f'+\frac{f}{r_c}\right)\cos\delta_0
+\mathcal{O}\!\left(\frac{r_c^2}{R^2}\right),
\label{eq:vrt}\\
\langle\Omega\rangle_\phi
=
\left(f'+\frac{f}{r_c}\right)\sin\delta_0
+\mathcal{O}\!\left(\frac{r_c^2}{R^2}\right).
\label{eq:div}
\end{align}
where $\delta_0=-\pi-\phi_0$ and $f$ and $f'$ are evaluated at $r=r_c$. Thus, to the order retained, the morphology measured by $\kappa$ is controlled solely by the local orientation $\delta_0$.

Projecting Eq.(\ref{eq:polarization}) perpendicular to the local polarization direction and performing the same angular average gives
\begin{align}
\left\langle\dot{\delta}\right\rangle_\phi
=
-Q_c\sin\delta_0
+\mathcal{O}\!\left(\frac{r_c^2}{R^2}\right),
\label{eq:deltadot}\\
Q_c=
\lambda_1\frac{f(r_c)}{r_c}
+\lambda_3 f'(r_c)>0.
\label{eq:qc}
\end{align}
Consequently,
$\dot{\kappa}=(d\kappa/d\delta_0)
\langle\dot{\delta}\rangle_\phi$.
Evaluating this relation for the five initial morphologies (see
\textcolor{red}{SM(sectionB5)}) yields
\begin{equation}
\dot{\kappa}_{\rm CA}=\dot{\kappa}_{\rm DA}=0,\qquad
\dot{\kappa}_{\rm CS}<0,\qquad
\dot{\kappa}_{\rm DS}>0,\qquad
\dot{\kappa}_{\rm VR}>0 .
\label{eq:kappa_dot}
\end{equation}
Eq.(\ref{eq:kappa_dot}) predicts that both asters initially retain their radial character, whereas the converging spiral becomes more vortex-like and the diverging spiral more
aster-like. The vortex also develops a radial component. Thus, the polarization-rate modes describe the initial changes in the local texture, while $\kappa$ characterizes
the resulting change in morphology.\\ 
These predictions apply only to the initial evolution. At later times, the density inhomogeneities generated by the different defect textures produce gradients that
feed back onto the polarization through Eq.(\ref{eq:polarization}).
The subsequent morphological evolution is therefore governed by the coupled nonlinear dynamics of density and polarization. As shown in Fig.\ref{fig:shape3.pdf}(a,d), both asters remain predominantly radial [see SM Movie 2.1 and Movie 2.4], while the converging spiral (Fig.\ref{fig:shape3.pdf}(b)) becomes
more vortex-like [see SM Movie 2.2] and the vortex develops radial character(Fig.\ref{fig:shape3.pdf}(c)) [see SM Movie 2.3]. The diverging spiral initially evolves toward larger $\kappa$ (Fig.\ref{fig:shape3.pdf}(e)), consistent with Eq. (\ref{eq:kappa_dot}) [see SM Movie 2.5], but its subsequent
evolution depends on activity and $\sigma$, and may even reverse direction. This behavior suggests a role for the evolving density--polarization coupling, alongside the other nonlinear terms.\\
Comparing Fig.~\ref{fig:shape3.pdf} with the defect mobility results in Fig.~\ref{fig:def_traj4.pdf}(f-j), reveals that stronger morphological evolution does not necessarily imply greater defect mobility. The two asters remain predominantly radial but exhibit
the strongest mobility asymmetries with opposite signs, consistent with their opposite longitudinal density-rate responses. The spirals undergo larger morphological changes but exhibit weaker mobility asymmetries, accompanied by reduced longitudinal and finite transverse
density responses. The vortex, with no initial
longitudinal density-rate mode, develops radial character but exhibits a weaker, parameter-dependent mobility asymmetry. These results indicate that defect motion depends not simply on the extent of morphological change, but on the density redistribution and polarization
dynamics associated with the initial texture.\\
\textit{Discussion and outlook.---}
Our results show that topological charge alone does not determine the annihilation dynamics of integer defects in dry polar active matter. At fixed pair geometry, changing the morphology of the $+1$ defect changes the relative motion of the pair; reversing the radial orientation of an aster reverses the mobility asymmetry, with the strongest response occurring for asters.\\
The early-time analysis links this morphology dependence to the coupled density and polarization dynamics. Different textures generate distinct density modes, which establish gradients that subsequently feed back on polarization. The initial polarization-rate modes constrain how each texture can deform, while the evolution of $\kappa$ tracks its changing radial and vortical character. The strong suppression of the mobility difference when density-gradient coupling is removed supports the importance of this feedback.\\
Integer defects also organize flows and mechanical structure in cellular active systems\cite{zhao2025integer}. Confined myoblast monolayers, for example, exhibit $+1$ spirals and asters with distinct cellular flows\cite{blanch2021integer,blanch2021quantifying}, and changes in cell density can drive a reversible spiral-to-aster transformation; such organizations have also been associated with localized stresses, cell differentiation, and three-dimensional tissue remodeling \cite{guillamat2022integer,dedenon2023density}. Our results suggest that morphology may additionally influence relative defect motion when such structures are free to move and interact. Whether this changes defect encounters, lifetimes, and annihilation pathways in unconfined cellular systems or many-defect polar active matter remains an open question.
\section{Acknowledgement}
The support and the resources provided by PARAM
Shivay Facility under the National Supercomputing Mision, Government of India at the Indian Institute of
Technology, Varanasi are gratefully acknowledged by all authors. S M thanks DST, SERB (INDIA), Project No.: CRG/2021/006945, MTR/2021/000438, and ANRF grant numbered ANRF/ARG/2025/008220/PS for financial support.

\bibliography{references}  

\begin{thebibliography}{4}%
\makeatletter
\providecommand \@ifxundefined [1]{%
 \@ifx{#1\undefined}
}%
\providecommand \@ifnum [1]{%
 \ifnum #1\expandafter \@firstoftwo
 \else \expandafter \@secondoftwo
 \fi
}%
\providecommand \@ifx [1]{%
 \ifx #1\expandafter \@firstoftwo
 \else \expandafter \@secondoftwo
 \fi
}%
\providecommand \natexlab [1]{#1}%
\providecommand \enquote  [1]{``#1''}%
\providecommand \bibnamefont  [1]{#1}%
\providecommand \bibfnamefont [1]{#1}%
\providecommand \citenamefont [1]{#1}%
\providecommand \href@noop [0]{\@secondoftwo}%
\providecommand \href [0]{\begingroup \@sanitize@url \@href}%
\providecommand \@href[1]{\@@startlink{#1}\@@href}%
\providecommand \@@href[1]{\endgroup#1\@@endlink}%
\providecommand \@sanitize@url [0]{\catcode `\\12\catcode `\$12\catcode `\&12\catcode `\#12\catcode `\^12\catcode `\_12\catcode `\%12\relax}%
\providecommand \@@startlink[1]{}%
\providecommand \@@endlink[0]{}%
\providecommand \url  [0]{\begingroup\@sanitize@url \@url }%
\providecommand \@url [1]{\endgroup\@href {#1}{\urlprefix }}%
\providecommand \urlprefix  [0]{URL }%
\providecommand \Eprint [0]{\href }%
\providecommand \doibase [0]{https://doi.org/}%
\providecommand \selectlanguage [0]{\@gobble}%
\providecommand \bibinfo  [0]{\@secondoftwo}%
\providecommand \bibfield  [0]{\@secondoftwo}%
\providecommand \translation [1]{[#1]}%
\providecommand \BibitemOpen [0]{}%
\providecommand \bibitemStop [0]{}%
\providecommand \bibitemNoStop [0]{.\EOS\space}%
\providecommand \EOS [0]{\spacefactor3000\relax}%
\providecommand \BibitemShut  [1]{\csname bibitem#1\endcsname}%
\let\auto@bib@innerbib\@empty
\bibitem [{\citenamefont {Harth}\ and\ \citenamefont {Stannarius}(2020)}]{harth2020topological}%
  \BibitemOpen
  \bibfield  {author} {\bibinfo {author} {\bibfnamefont {K.}~\bibnamefont {Harth}}\ and\ \bibinfo {author} {\bibfnamefont {R.}~\bibnamefont {Stannarius}},\ }\bibfield  {title} {\bibinfo {title} {Topological point defects of liquid crystals in quasi-two-dimensional geometries},\ }\href@noop {} {\bibfield  {journal} {\bibinfo  {journal} {Frontiers in Physics}\ }\textbf {\bibinfo {volume} {8}},\ \bibinfo {pages} {112} (\bibinfo {year} {2020})}\BibitemShut {NoStop}%
\bibitem [{\citenamefont {Missaoui}\ \emph {et~al.}(2020)\citenamefont {Missaoui}, \citenamefont {Harth}, \citenamefont {Salamon},\ and\ \citenamefont {Stannarius}}]{missaoui2020annihilation}%
  \BibitemOpen
  \bibfield  {author} {\bibinfo {author} {\bibfnamefont {A.}~\bibnamefont {Missaoui}}, \bibinfo {author} {\bibfnamefont {K.}~\bibnamefont {Harth}}, \bibinfo {author} {\bibfnamefont {P.}~\bibnamefont {Salamon}},\ and\ \bibinfo {author} {\bibfnamefont {R.}~\bibnamefont {Stannarius}},\ }\bibfield  {title} {\bibinfo {title} {Annihilation of point defect pairs in freely suspended liquid-crystal films},\ }\href@noop {} {\bibfield  {journal} {\bibinfo  {journal} {Physical Review Research}\ }\textbf {\bibinfo {volume} {2}},\ \bibinfo {pages} {013080} (\bibinfo {year} {2020})}\BibitemShut {NoStop}%
\bibitem [{\citenamefont {Kruse}\ \emph {et~al.}(2004)\citenamefont {Kruse}, \citenamefont {Joanny}, \citenamefont {J{\"u}licher}, \citenamefont {Prost},\ and\ \citenamefont {Sekimoto}}]{kruse2004asters}%
  \BibitemOpen
  \bibfield  {author} {\bibinfo {author} {\bibfnamefont {K.}~\bibnamefont {Kruse}}, \bibinfo {author} {\bibfnamefont {J.~F.}\ \bibnamefont {Joanny}}, \bibinfo {author} {\bibfnamefont {F.}~\bibnamefont {J{\"u}licher}}, \bibinfo {author} {\bibfnamefont {J.}~\bibnamefont {Prost}},\ and\ \bibinfo {author} {\bibfnamefont {K.}~\bibnamefont {Sekimoto}},\ }\bibfield  {title} {\bibinfo {title} {Asters, vortices, and rotating spirals in active gels of polar filaments},\ }\href {https://doi.org/10.1103/PhysRevLett.92.078101} {\bibfield  {journal} {\bibinfo  {journal} {Physical Review Letters}\ }\textbf {\bibinfo {volume} {92}},\ \bibinfo {pages} {078101} (\bibinfo {year} {2004})}\BibitemShut {NoStop}%
\bibitem [{\citenamefont {R{\o}nning}\ \emph {et~al.}(2023)\citenamefont {R{\o}nning}, \citenamefont {Renaud}, \citenamefont {Doostmohammadi},\ and\ \citenamefont {Angheluta}}]{ronning2023spontaneous}%
  \BibitemOpen
  \bibfield  {author} {\bibinfo {author} {\bibfnamefont {J.}~\bibnamefont {R{\o}nning}}, \bibinfo {author} {\bibfnamefont {J.}~\bibnamefont {Renaud}}, \bibinfo {author} {\bibfnamefont {A.}~\bibnamefont {Doostmohammadi}},\ and\ \bibinfo {author} {\bibfnamefont {L.}~\bibnamefont {Angheluta}},\ }\bibfield  {title} {\bibinfo {title} {Spontaneous flows and dynamics of full-integer topological defects in polar active matter},\ }\href {https://doi.org/10.1039/D3SM00316G} {\bibfield  {journal} {\bibinfo  {journal} {Soft Matter}\ }\textbf {\bibinfo {volume} {19}},\ \bibinfo {pages} {7513} (\bibinfo {year} {2023})}\BibitemShut {NoStop}%
\end{thebibliography}%

\end{document}